\documentstyle[epsf]{mn}

\begin{document}
\def\lsun{{\rm L_{\odot}}}
\def\msun{{\rm M_{\odot}}}
\def\rsun{{\rm R_{\odot}}}
\title{The Population of Faint Transients in the Galactic Centre}
\author[A.R. King]{A.R. King$^1$\\ 
$^1$ Astronomy Group, University of Leicester, Leicester LE1
7RH, U.K}
\maketitle            
\begin{abstract}
BeppoSAX has detected a population of faint transient X--ray sources in
the Galactic Centre.
I show that a simple irradiated--disc picture gives a consistent fit
to the properties of this population, and that
it probably consists of low--mass X--ray binaries which
have evolved beyond their minimum orbital periods $\sim 80$~min.
Since all post--minimum systems are transient, and neutron--star LMXBs
are more common than black--hole LMXBs in the Galaxy, the
majority of these systems should contain neutron stars, as observed.  This
picture predicts that the Galactic Centre transients should have
orbital periods in the range $\sim 80 - 120$~minutes, and that most of
them should repeat in the next few years. In this case, the total
number of post--minimum transients in the Galaxy would be considerably
smaller than the usual estimates of its total LMXB population. I
discuss possible reasons for this.

\end{abstract}

\section{Introduction}
Over the past 3~yr the Wide Field Cameras on BeppoSAX have detected
9 soft X--ray transients in a 40x40 degree field around the Galactic
Centre (GC) (see Heise et al. 1998 for a review).
These outbursts are rather faint ($L_X \la 3\times 10^{37}$~erg~s$^{-1}$)
and have very short e--folding timescales $\tau \sim 2-6$~d. Of the 9
detected systems, 7 have shown Type~I X--ray bursts, indicating that
the accretor is a neutron star. This statistic contrasts sharply with
that for brighter transients in the Galaxy, which suggests that no
more than one half contain neutron stars, while a large fraction are
believed on the basis of dynamical mass measurements to contain black
holes.

The faintness and short decay timescales of the GC transient outbursts
imply a remarkably low accreted mass
\begin{equation}
\Delta M \simeq 1.8\times 10^{-11}L_{37}\tau_4\msun,
\label{m}
\end{equation}
where $L_{37}, \tau_4$ are $L_x, \tau$ in units of
$10^{37}$~erg~s$^{-1}$ and 4~d respectively, and we have assumed that
each gram of accreted matter releases $10^{20}$~erg. (This mass
estimate would rise if a significant fraction of the accreted matter
radiated at low efficiency, i.e. was advected. This is only possible
for black--hole systems.)  The limits quoted above show that 
the maximum value of $L_{37}\tau_4$ is about 4.5, allowing $\Delta M$ as large
as $8\times 10^{-11}\msun$.

Only two of the GC transients, SAX
J1808.4-3658, and very recently SAX J1747.0-2853 (Markwardt et
al. 2000) have been observed to recur. If all of the mass transferred
from the companion is accreted efficiently by the compact primary, one
can then estimate its mass transfer rate in SAX J1808.4-3658 as $\dot
M_{\rm tr} \simeq 1\times 10^{-11}\ \msun\ {\rm yr}^{-1}$ (Chakrabarty
\& Morgan, 1998). The non--recurrence of the other GC transients
implies that their mass transfer rates must also be very low, i.e.
\begin{equation}
\dot M_{\rm tr} \la \Delta M/({\rm 3~yr}) 
\sim 6\times 10^{-12}L_{37}\tau_4\ \msun{\rm yr}^{-1}.
\label{mdot}
\end{equation}
We shall see in Section 2 below that explicit modelling of the
accretion disc stability in these systems leads to a similar
limit. (Note that the estimate (\ref{mdot}) would no longer be an upper
limit if, contrary to the assumption made above, a significant
fraction of the transferred mass was lost from the system rather than
accreted, cf Meyer \& Meyer--Hofmeister, 1994).

There seem to be only two kinds of relatively abundant binary
system which would have such low transfer rates:

1. wind--fed systems\\

2. Roche--lobe--filling systems in which the donor star has an
extremely low mass $M_2 < 0.1\msun$, with mass transfer driven by
gravitational radiation. This gives a rate (cf King, Kolb \&
Szuszkiewicz, 1997)
\begin{equation}
   \dot M_{\rm GR} \simeq 1 \times 10^{-11}\:
  m_1^{2/3}\biggl({m_2\over 0.1}\biggr)^2\: \biggl({P\over 2\ {\rm
  hr}}\biggr)^{-8/3} \: \msun {\rm yr}^{-1} \, , \label{gr}
\end{equation}
where $m_1, m_2$ are the accretor and donor masses in $\msun$, and I
have assumed $m_1 >> m_2$.

Note that 1. need not automatically imply a high--mass donor, which
would be in possible contradiction with the usual identification of
soft X--ray transients as low--mass X--ray binaries (LMXBs). For the
mass transfer rates (\ref{mdot}) are so low that even the
comparatively weak wind of a low--mass main sequence star may be
sufficient to power them. Alternatively, radio pulsar irradiation
could excite a wind from a low--mass companion star. Orbital angular
momentum losses may then shrink the binary to the point where
the neutron star begins to capture this wind material and turns on as an
X--ray source. The source will be transient because the accretion rate
through the disc is too low to keep it fully ionized (see equation
\ref{mtr} below).

A possible example of a higher--mass wind--fed transient may be CI Cam
(e.g. Marshall et al., 1998, Clark et al., 2000), which has been
variously classified as a symbiotic system, or a B[e] binary. Any such
system must be wide enough that the UV radiation from the high--mass
donor is unable to keep the accretion disc ionized.  Whatever the mass
of the donor, accretion by wind capture will tend to produce an
accretion disc of small radius: if much of the disc mass is accreted
in an outburst, the outer radius of the disc will be determined by the
low specific angular momentum of the captured wind material, naturally
producing a small disc. This in turn would account for the low
accreted mass $\Delta M$ and short decay times $\tau$ (see equations
\ref{dm}, \ref{tth} below).  One might test for the presence of
high--mass companions of this type in the GC transients through
optical/IR identifications. There are currently only two reported
identifications, both of which are inconsistent with high--mass
stars. In SAX J1808.4-3658 (Roche et al., 1998) the 2--hour orbital
period rules out any such companion. In SAX J1810.8-2609, Greiner et
al. (1999) find $R = 19$ fading to $R > 21.5$ for the optical/IR
source. This variation means that this source is presumably dominated
by the accretion disc, which is probably irradiated by the central
X--rays. This is inconsistent with a high--mass companion, which would
dominate the optical/infrared as it would
have a similar effective temperature to the disc, but be much larger.

These arguments fall short of proving that {\it no} GC transient is
wind--fed, in either high--mass or low--mass versions. However the
second possibility listed above seems inescapable. At the end of their
lives as accreting binaries, LMXBs must reduce the donor mass $M_2$ to
the point where this star begins to become degenerate (i.e. $M_2 \la
0.1\msun$). If the orbital evolution of LMXBs is similar to the
standard picture assumed for cataclysmic variables (CVs), where the
accretor is a white dwarf rather than a neutron star or black hole,
the binary period will be close to $\sim 80$~minutes at this point
(see e.g. King, 1988; Kolb \& Baraffe, 1999 for reviews). Subsequently
this period begins to increase rather than decrease as orbital angular
momentum is lost via gravitational radiation, as the donor expands in
response to further mass loss. The mass transfer rate drops to values
comparable with those of equation (\ref{mdot}). Because of this, the
mass transfer timescale $M_2/\dot M_{\rm tr}$ becomes very long,
eventually approaching a Hubble time. The system's evolution thus
slows considerably; even a system `born' at the minimum period of
$\sim 80$~minutes requires a time of order the age of the Galaxy to
reach $P\simeq 2$~hr. Mass transfer continues, albeit at a very slow
rate ($\sim (0.5 - 1)\times 10^{-11}\ \msun\ {\rm yr}^{-1}$). Hence
there should exist a population of such extremely faint LMXB systems,
directly analogous to the post--minimum--period population of
CVs. While no post--minimum CV has been certainly identified, their
LMXB analogues are potentially much easier to observe since they are
all likely to be soft X--ray transients (King, Kolb \& Szuszkiewicz,
1997). As we shall see, their expected properties are very similar to
those of the faint Galactic Centre transients. Note that there may be
other more exotic ways of producing individual systems among the faint
GC transients, as has for example been proposed for SAX J1808.4-3658
by Ergma \& Antipova (1999). However post--minimum systems offer the
most likely way of producing them in significant numbers.

The identification of the faint GC transients with post--minimum LMXBs
has the desirable property of explaining why neutron stars are
favoured among this population.  Black--hole LMXBs are likely to be
transient at all orbital periods (King, Kolb \& Szuszkiewicz, 1997),
while neutron--star LMXBs with main--sequence donors (i.e. before they
reach the minimum period) are likely to be persistent. Accordingly
black--hole systems predominate among bright transients, since a
bright neutron--star transient requires an unusual (nuclear--evolved)
donor. However the calculations of King, Kolb \& Szuszkiewicz (1997)
show that neutron--star LMXBs {\it will} become transient as they
evolve beyond the the minimum period. Thus the ratio of neutron star
to black hole systems must be much higher for post--minimum LMXBs than
for bright transients. This naturally explains why the faint Galactic
Centre transients seem mostly to contain neutron stars.

In the next Section I show that the simple irradiated--disc model of
soft X--ray transient (SXT) outbursts proposed by King \& Ritter
(1998) predicts outburst masses $\Delta M$ and decay times $\tau$ in
excellent agreement with the observational estimates given above,
provided that the outer disc radius is small, i.e. $\sim 10^{10}$~cm.
Moreover the maximum average mass transfer rate consistent with
irradiation not suppressing the outbursts is close to the limit
(\ref{mdot}). In Section 3 I discuss the two possible types of binary
system considered above in the light of these disc properties. Section
4 is the Conclusion.

\section{Accretion disc properties in the faint GC transients}

Soft X--ray transient outbursts are thought to result from
instabilities in LMXB accretion discs. Both the incidence and the
nature of these outbursts are strongly affected by irradiation of the
disc surfaces by the central X--rays. Irradiation can suppress
outbursts in some LMXB discs by removing their hydrogen ionization
zones (van Paradijs, 1996; King, Kolb \& Burderi, 1996; King, Kolb \&
Szuszkiewicz, 1997), thus making them stable (persistent) at lower
mass transfer rates than is true for the otherwise similar discs in
CVs. The irradiation effect appears to be weaker if the accretor is a
black hole rather than a neutron star, possibly because of the lack of
a hard surface (King, Kolb \& Szuszkiewicz, 1997). The result is that
neutron--star LMXBs with main--sequence companions tend to be
persistent, while similar black--hole binaries are largely transient.

If an LMXB disc goes into outburst, irradiation of the disc greatly
prolongs the high state, and causes viscous rather than thermal
effects to dominate the light--curve (King \& Ritter, 1998). This
often produces an exponential decay, particularly if the whole disc is
efficiently irradiated. Several predictions of this simple
irradiated--disc picture are confirmed by observation (cf Shahbaz,
Charles \& King, 1998). More detailed calculations with a full 1--D
disc code (Dubus et al, 1999) give results very similar to those of
King and Ritter (1998) so I shall use their simple analytic
expressions in the following.

As the observed decays are approximately exponential I assume that
the entire disc of a faint transient (of radius $R$) is irradiated
during an outburst. Most of this mass is accreted.  Immediately
before the outburst the surface density in the disc must have been
close to the value
\begin{equation}
\Sigma_{\rm max} = 11.4R_{10}^{1.05}m_1^{-0.35}\alpha_c^{-0.86}\ {\rm
g\ cm}^{-2}
\end{equation}
(Cannizzo, Shafter \& Wheeler, 1988)
triggering the thermal instability through local viscous dissipation,
where $R_{10} = R/10^{10}$~cm, with $R$ the radial disc coordinate,
$m_1$ is the central accreting mass in $\msun$, and $\alpha_c$ is the
cold--state viscosity parameter. Thus by integrating over $R$ we predict
the mass accreted in the outburst as
\begin{equation}
\Delta M_{\rm pr} \simeq 1.5\times 10^{-11}
m_1^{-0.35}\alpha_{0.05}^{-0.86}R_{10}^{3.05}\msun,
\label{dm}
\end{equation}
where $\alpha_{0.05} =
\alpha_c/0.05$. This relation can be compared with equation (8) of
King \& Ritter (1998), who used a simpler form of $\Sigma_{\rm max}$.
Equating this to the observational estimate (\ref{m}) we find a
disc radius 
\begin{equation}
R \simeq 1.2\times 10^{10}
(L_{37}\tau_4)^{0.33}m_1^{0.11}\alpha_{0.05}^{0.28}\ {\rm cm}.
\label{R}
\end{equation}
King \& Ritter (1998) predict that the e--folding time for the decay will
be 
\begin{equation}
\tau_{\rm pr} = {R^2\over 3\nu},
\end{equation}
where $\nu = \alpha_hc_SH$ is the hot--state kinematic viscosity at
the disc edge. Here $c_S, H$ are the local sound speed and scale
height, and we have
\begin{equation}
H = {c_S\over \Omega}
\label{H}
\end{equation}
with $\Omega = (GM_1/R^3)^{1/2}$ the Kepler frequency at disc radius $R$.
Writing $T_4$ for the surface temperature at this point in units of
$10^4$~K, we find the predicted timescale
\begin{equation}
\tau_{\rm pr} = 3.9\alpha_h^{-1}m_1^{0.5}R_{10}^{0.5}T_4^{-1}\ {\rm d}.
\label{tth}
\end{equation}

The theoretical predictions (\ref{dm}, \ref{tth}) are in
excellent agreement with the observational values $\Delta M$, $\tau$
and thus $L_X$ provided that the hot--state viscosity parameter takes
a value $\alpha_h \sim T_4 \la 1$, as expected, and $R \sim
10^{10}$~cm. Evidently the distinctive feature of the discs in the
faint GC transients is their very small size, which accounts for both
their low peak luminosities and their rapid decays. We can now ask
how low the mean mass transfer rate $\dot M_{\rm tr}$ must be if such
a disc is to have ionization zones despite being so small. This
requires that the surface temperature $T_{\rm irr}$ resulting from
irradiation should be less than the ionization temperature $T_{\rm H}
\sim 6500$~K at the disc edge. I take
\begin{equation}
T_{\rm irr}(R)^4 = {10^{20}\dot M_{\rm tr}\over 4\pi \sigma R^2}
\biggl({H \over R}\biggr)^n
\biggl[{{\rm d}\ln H\over {\rm d}\ln R} - 1\biggr], 
\label{eq3}
\end{equation}
(e.g. van Paradijs, 1996) where the factor in square 
brackets lies between 1/8 and 2/7 and the index $n = 1$ or 2 for a
neutron star or black hole respectively (cf King, Kolb \&
Szuszkiewicz, 1997). Requiring $T_{\rm irr} < T_{\rm H}$ and using
the estimate (\ref{R}) gives
\begin{equation}
\dot M_{\rm tr} \la 1.3\times 10^{-11}
(L_{37}\tau_4)^{0.66}m_1^{0.22}\alpha_{0.05}^{0.56}\ \msun {\rm yr}^{-1}
\label{mtr}
\end{equation}
for a neutron star, and a limit about 10 times larger for a black
hole. This is again in excellent agreement with the observational 
limit~(\ref{mdot}). 

\section{Binary models for the faint GC transients}

The good agreement with simple irradiated--disc theory found above
shows that the faint GC transient population must be binaries with two
key properties

(a) disc radii $R \sim 1.2\times 10^{10}$~cm, and

(b) mean mass transfer rates 
$\dot M_{\rm tr} \sim (0.6 - 1)\times 10^{-11}\ \msun{\rm yr}^{-1}$.

Both the binary types 1 (wind--fed) and 2 (low--mass donor)
considered in the Introduction are able to reproduce these
conditions. In the wind--fed case, the low specific angular momentum
of matter captured from a wind will produce a small disc radius
provided that most of the disc mass is accreted during an
outburst. The low mass transfer rate is a natural consequence of
inefficient wind capture. However, the specific values of $R$ and
$\dot M_{\rm tr}$ depend in detail on the properties of the donor
wind, and it is not obvious why these should cluster around the values
producing (a) and (b) above. By contrast, the version of model 2
involving post--minimum LMXBs will naturally produce these values.  If
most of the disc mass is accreted in an outburst, as is implicit in
the estimate (\ref{dm}), the disc radius $R$ will be close to the
circularization radius 
\begin{equation}
R_{\rm circ} \simeq (1+q)(0.7-0.227\log q)^4a,
\label{circ}
\end{equation}
(e.g. Frank et al, 1992) where $q = M_2/M_1$ is the mass ratio and $a$ is
the binary separation, i.e.
\begin{equation}
a = 3.53\times 10^{10}m_1^{1/3}P_{\rm hr}^{2/3}\ {\rm cm}.
\label{a}
\end{equation}
With $M_1 = 1.4\msun, M_2 = 0.1\msun$ we find an expected disc radius
\begin{equation}
R \simeq R_{\rm circ} = 1.7\times 10^{10}\biggl({m_1\over 1.4}\biggr)^{1/3}
\biggl({P\over 80\ {\rm min}}\biggr)^{2/3}\ {\rm cm}.
\label{rcirc}
\end{equation}
Thus binary periods in the typical range 80~min -- 2~hr will produce
discs of the right size (cf eqn. \ref{R}) for neutron--star masses
$m_1 = 1.4$. Equating the two expressions (\ref{R}, \ref{rcirc}) gives
the requirement
\begin{equation}
L_{37}\tau_4 = 5m_1^{2/3}\alpha_{0.05}^{-0.84}
\biggl({P\over 2\ {\rm hr}}\biggr)^2,
\label{Lt}
\end{equation} 
which compares well with the observed range of this quantity given in
the Introduction. The calculations of King, Kolb \& Szuszkiewicz
(1997) show that such systems have transfer rates at or below the
limit (b), which also ensure that the systems are indeed transient
there (see their Fig. 3).

For black--hole masses $m_1 \sim 7$ we see that the predicted disc
radius $R$ tends to become rather larger than the estimate (\ref{R}),
leading to more prolonged outbursts (transferred mass $\Delta M_{\rm
pr} \sim 10^{-10}\msun$, decay timescale $\tau_{\rm pr} \sim 27$~d)
than are typical of the faint GC transients (note that eqs. \ref{dm},
\ref{rcirc} imply that $\Delta M_{\rm pr} \propto m_1$, while
eq. \ref{tth} implies $\tau_{\rm pr} \propto m_1$). Thus such systems
would probably not be classified as faint transients. Moreover
black--hole systems will have longer recurrence times $t_{\rm rep}
\propto \Delta M/\dot M_{\rm tr} \propto m_1/m_1^{2/3} \sim m_1^{1/3}$
and thus lower discovery probability ($\dot M_{\rm tr} \propto
m_1^{2/3}$ for gravitational radiation, cf eqn \ref{gr}). This already
tends to suggest agreement with the observation that at least 7 out of
9 GC transients contain neutron stars. A still stronger reason comes
from the fact that Fig. 3 of King, Kolb \& Szuszkiewicz (1997) shows
that essentially {\it all} neutron--star LMXBs will be transient once
they have evolved sufficiently far beyond the minimum period, and
neutron--star systems are much more common than black--hole ones in
general. Thus the ratio of neutron--star to black--hole systems among
the faint GC transients should be
\begin{eqnarray}
\lefteqn{{N_{\rm NS\ GC transients}\over N_{\rm BH\ GC transients}} 
\simeq {N_{\rm NS\ post}\over N_{\rm BH\ post}}} \nonumber \\ 
\lefteqn{= {N_{\rm NS\ post}\over N_{\rm NS\ pre}}.
{N_{\rm NS\ pre}\over N_{\rm BH\ pre}}.
{N_{\rm BH\ pre}\over N_{\rm BH\ post}}},
\label{N}
\end{eqnarray}
where the $N$ are space densities and `pre' and `post' refer to the
minimum period. Now Fig. 3 of King, Kolb \& Szuszkiewicz
(1997) implies
\begin{equation}
{N_{\rm NS\ post}\over N_{\rm NS\ pre}} \ga 
{N_{\rm BH\ post}\over N_{\rm BH\ pre}},
\label{N2}
\end{equation}
i.e. the slow--down of neutron--star LMXB evolution after passing the
minimum period is more dramatic than that of black--hole systems. Using
(\ref{N2}) in (\ref{N}), the two outer factors combine to give a
number $\ga 1$, so we get
\begin{equation}
{N_{\rm NS\ GC transients}\over N_{\rm BH\ GC transients}}
\ga {N_{\rm NS\ pre}\over N_{\rm BH\ pre}} >> 1,
\label{N4}
\end{equation}
where the last inequality expresses the fact that there are far more
neutron--star then black--hole binaries.

\section{Conclusions}

I have shown that the simple irradiated--disc picture gives a
consistent fit to the properties of the faint GC transients, and that
this population probably consists of post--minimum LMXBs. Neutron star
systems far outweigh black--hole ones here, mainly because {\it all}
post--minimum systems are transient, and neutron--star LMXBs are
simply more common than black--hole ones in the Galaxy (outbursts of
the latter would also probably be too long to be classified among the
faint transients). By contrast among brighter transients, which are
generally pre--minimum systems, the greater incidence of transient
behaviour among black--hole systems makes them prominent. If the
identification as post--minimum LMXBs is correct, the faint transients
should be binaries with periods in the range $80 - 120$~minutes, with
extremely low--mass companions. 

An interesting point emerges from the fact that only two faint GC
transients have yet been observed to repeat. From (\ref{mdot}) we see
that if this state of affairs persists for a decade or so, the
resulting upper limit on the mean mass transfer rate will become
embarassingly low even for post--minimum LMXBs. On the other hand, a
typical repetition time $t_{\rm rep}$ of order a few years would also
severely limit the total number of such systems in the Galaxy: Heise
et al. (1998) estimate from the sky and temporal coverage of the
BeppoSAX WFC that there are about 18 faint transient outbursts in the
Galaxy per year, leading to a total population $N_{\rm Gal} \sim
18(t_{\rm rep}/{\rm yr})$. This number ($50 \la N_{\rm Gal} \la 180$)
is considerably smaller than the usual estimate of $\sim 1000$ LMXBs
in the Galaxy, whereas one might expect it to be much larger, as the
evolution of post--minimum systems is so slow. There are several
possible reasons for this, of which two seem most likely.

(a) If most LMXBs first reach contact at initial periods $P_i$ greater
than a few hours (as indeed suggested by theoretical studies of LMXB
formation, e.g. Kalogera \& Webbink, 1996; King \& Kolb 1997), their
lifetimes before becoming post--minimum transients may be comparable
with the age of the Galaxy. This lifetime is spent mostly near the
minimum period (cf Kolb \& Baraffe, 1999) while according to King,
Kolb \& Szuszkiewicz (1997), neutron--star LMXBs have to evolve
somewhat beyond this period in order to become transient. A large
fraction of the neutron--star LMXBs ever formed in the Galaxy may
still not have reached this stage.

(b) Isolated millisecond pulsars are thought to be neutron stars spun
up in LMXBs, but which have evaporated their companions (see
Bhattacharya \& van den Heuvel, 1991, for a review). This suggests
that many neutron--star LMXBs may not even reach the theoretical
minimum period, and thus do not become faint transients at all. It is
sometimes hypothesized (cf Bhattacharya \& van den Heuvel, 1991) that
most neutron--star LMXBs evaporate the companion star as the system
attempts to cross the analogue of the CV period gap between 3~hr and
2~hr (see e.g. King, 1988 for a review of the latter). If so, this
would limit post--minimum neutron--star systems to the rare examples
first coming into contact at periods below the period gap, i.e. with
$P_i \la 2$~hr. Of course there can be no such effect for black--hole
LMXBs.

At present we do not have any clear idea of the mean repetition time
for the faint GC transients. The arguments above show that there are
interesting consequences whatever this number turns out to be.
Extensive X--ray monitoring of the Galactic Centre region clearly has
much more to tell us about the stellar populations there.

\section{Acknowledgment}
I gratefully acknowledge the support of a PPARC Senior Fellowship. I
thank John Heise, Uli Kolb, Hans Ritter and Klaus Schenker for useful
discussions, and the referee for a perceptive and helpful report.

\end{document}